\documentclass[a4paper]{article}

\usepackage{INTERSPEECH2020}
\usepackage{xcolor}
\usepackage{hyperref}
\usepackage[colorinlistoftodos,prependcaption,textsize=footnotesize]{todonotes}
\usepackage{tikz}
\usepackage{pgfplots}
\usepgfplotslibrary{groupplots}
\usepackage{chngpage}
\usepackage{soul}

\title{Self-Supervised Representations Improve End-to-End Speech Translation}
\name{Anne Wu, Changhan Wang, Juan Pino, Jiatao Gu}
\address{Facebook AI, USA}
\email{\{annewu, changhan, juancarabina, jgu\}@fb.com}

\begin{document}

\maketitle

\begin{abstract}

End-to-end speech-to-text translation can provide a simpler and smaller system but is facing the challenge of data scarcity. 
Pre-training methods can leverage unlabeled data and have been shown to be effective on data-scarce settings. 
In this work, we explore whether self-supervised pre-trained speech representations can benefit the speech translation task in both high- and low-resource settings, whether they can transfer well to other languages, and whether they can be effectively combined with other common methods that help improve low-resource end-to-end speech translation such as using a pre-trained high-resource speech recognition system.
We demonstrate that self-supervised pre-trained features can consistently improve the translation performance, and cross-lingual transfer allows to extend to a variety of languages without or with little tuning.

\end{abstract}

\section{Introduction}

Recently, there has been much interest in end-to-end speech translation (ST) models \cite{berard2016listen, duong2016attentional, weiss2017sequence, berard2018end, vila2018end, di2019adapting, inaguma2019multilingual}, which, compared to traditional cascaded models, are simpler and computationally more efficient, can preserve more acoustic information and can avoid propagating errors from the speech recognition component. Large amounts of annotated data are usually required for achieving a good performance for such systems, but supervised training data for ST remain very limited.

On the other hand, unlabeled data are more accessible. Self-supervised techniques can exploit unlabeled data by learning a representation through, for instance, partial prediction or contrastive methods, and they have been shown effective for natural language \cite{devlin2018bert, radford2018improving, baevski2019cloze} and speech processing \cite{oord2018representation, schneider2019wav2vec, baevski2019vq}. In the latter case, several investigations on unsupervised or self-supervised pre-training have been conducted and applied to English automatic speech recognition (ASR) \cite{schneider2019wav2vec, baevski2019vq}, to multilingual ASR by training multilingual features \cite{kawakami2020learning} or transferring contrastive predictive coding (CPC) features to other languages \cite{riviere2020unsupervised}. 

In this paper, we are interested in whether self-supervised speech pre-training can effectively help speech-to-text translation on both high-resource and low-resource settings. In particular, we focus on the method of wav2vec \cite{schneider2019wav2vec} which makes use of contrastive predictive coding (CPC), the vector-quantized representation vq-wav2vec \cite{baevski2019vq} and BERT features learned on top of the discretized representations \cite{baevski2019vq}. 

We use speech features pre-trained on English, and first examine a high-resource within-language English-to-X ST setting (X denotes a non-English language), then we transfer the representations to 11 lower-resource X-to-English ST tasks. Transferring the parameters learned on a higher-resource ASR task has been shown to be an effective way to improve the performance and ameliorate the training of low-resource ST \cite{bansal2018pre, stoian2019analyzing, wang2019bridging}, thus we also study the interactions with self-supervised representations and whether we can effectively combine both methods. 

We first demonstrate that compared to commonly used log-mel filterbank features, self-supervised features pre-trained on English can help improve English-to-X ST, but also transfer well to other languages even without requiring additional tuning. However, in the cross-lingual case, training data quantity and linguistic similarity may affect this gain. Further study shows that either fine-tuning the pre-trained input features or using a multilingual ASR model to fine-tune the final ST system can both improve the X-to-English ST. Finally, we show that when using an ASR model to pre-train ST systems, under certain training conditions, the ASR performance may not be a good indicator of the ST performance.

\section{Methods}\label{sec:methods}

\subsection{Self-supervised Learning for Speech Representations}\label{sec:ssup}

Self-supervised learning allows to learn representations \cite{devlin2018bert, chen2020simple, oord2018representation, peters2018deep, lewis2019bart} through proxy tasks by, for instance, predicting some masked parts of the input, predicting future time-steps, contrasting with negative samples, or generating contextual data. In our case, we focus on three speech feature pre-training techniques which either makes use of CPC or a masked language model. 

In this work, we explore \textbf{four} self-supervised approaches for learning speech representations in ST.
The first and simplest representation is \textit{wav2vec}\cite{schneider2019wav2vec}, which learns speech representations through a future sample prediction task by optimizing a contrastive loss. The model consists of two convolutional neural networks, with an encoder network that takes raw audio as inputs and outputs a low-frequency representation to an aggregator, that creates a contextualized vector representation by combining the latent representation from multiple time steps. 
As a follow-up, \textit{vq-wav2vec}\cite{baevski2019vq} has an architecture similar to \textit{wav2vec}, but with an additional quantization module between the encoder network and the aggregator, which discretizes the encoder's outputs before feeding them to the aggregator network. The output representation, as discrete tokens, can be consumed by natural language processing algorithms/models such as BERT from which we can extract representations for speech tasks. We also investigate an approach leveraging the pre-trained BERT, described in \autoref{sec:network}.

\subsection{Network architecture}\label{sec:network}

For both ST and ASR tasks, our experiments are performed with a sequence-to-sequence BiLSTM attention-based encoder-decoder architecture following \cite{berard2018end}, but with a 3-layers decoder. Speech features are given as inputs to two non-linear $(tanh)$ layers, then passed to a stack of two convolutional layers. The output tensor is flattened and fed into three stacked bidirectional LSTM layers. The decoder is composed of two LSTM layers which output to a linear projection layer. 

For low-resource ST settings, we also investigate a hybrid BERT-backbone architecture, where we reuse the BERT model pre-trained on discretized speech features as the encoder. For the decoder, we keep the same architecture than the BiLSTM. While BERT is commonly used on monolingual tasks since it has been developed at first for natural language understanding, this allows to reuse it for a different goal and avoiding training an important number of parameters from scratch.

\section{Experiments}

\subsection{Datasets}

For English-to-X ST, we use the MuST-C~\cite{di2019must} dataset, a corpus with audio recordings from English TED talks translated into 8 languages. The corpus comprises sentence-level aligned transcriptions and translations.

For X-to-English ST, we use the multilingual ST dataset CoVoST \cite{wang2020covost} from 11 languages (French, German, Dutch, Russian, Spanish, Italian, Turkish, Persian, Swedish, Mongolian and Chinese) to English, containing crowd-sourced speech with diverse speakers and accents on a variety of topics, from dialogue to movie scripts. For ASR, we use the English data from the corresponding Common Voice dataset (2019-06-12 release), with approximately 120 hours \cite{ardila2019common}.
For the test set, we use the CoVoST test set for all the languages, and on the Tatoeba test set whenever it is available (i.e. for Fr, De, Nl, Ru and Es-En ST). Dataset statistics can be found in \autoref{tab:asr_data}.


\begin{table}[t]
  \caption{AST training data statistics. We also use the source language transcripts as the training data for ASR (if used).}
  \label{tab:asr_data}
  \centering
  \begin{tabular}{ccc|ccc}
       \toprule
       Pairs & Hours & Data & Pairs & Hours & Data \\
       \midrule
       Fr-En & 87h & CoVoST & Fa-En & 20h & CoVoST\\
       De-En & 71h & CoVoST & Sv-En & 1h & CoVoST\\
       Es-En & 21h & CoVoST & Mn-En & 3h & CoVoST\\
       Nl-En & 4h  & CoVoST & Zh-En & 4h & CoVoST\\
       Ru-En & 10h & CoVoST & \\
       It-En & 13h & CoVoST & En-Fr & 492h & MuST-C\\
       Tr-En & 3h & CoVoST & En-Ro & 432h & MuST-C\\
       \bottomrule
  \end{tabular}
  \vspace{-10pt}
\end{table}



\subsection{Self-supervised Pre-trained Models}
In our experiments, we use the officially open-sourced wav2vec \cite{schneider2019wav2vec}, vq-wav2vec (k-means) \cite{baevski2019vq} and BERT models \cite{baevski2019vq} \footnote{These models are available for download at \url{https://github.com/pytorch/fairseq/tree/master/examples/wav2vec}} trained on the full 960h of Librispeech corpora~\cite{panayotov2015librispeech}.

\subsection{Experimental Setups}
\subsubsection{Pipelines}
\label{sec:pipeline} 
For both high-resource and low-resource ST settings, we compute the log-mel filterbank features and extract the frozen learned features for direct ST training. For low-resource ST, we additionally pre-train an English ASR model with the corresponding speech features, then transfer the encoder or both the encoder and decoder parameters for warming-up ST training.

\subsubsection{Preprocessing}
For the preparation of transcript and translation, we normalize the punctuation, tokenize the text with sacreMoses and lowercase to align with previous settings \cite{wang2020covost} \cite{di2019must}. We remove the punctuations only from the transcripts. On CoVoST, we use a character-level vocabulary, with 54 characters including English alphabet and numerical characters, punctuations and the markers for fairseq \cite{ott2019fairseq} dictionary. On MuST-C, we choose a unigram vocabulary of size 10000 as in \cite{mccarthy2020skinaugment} to better balance the training time, as the sentences in MuST-C are generally longer. The vocabulary is obtained using SentencePiece \cite{kudo2018sentencepiece}.

We convert the raw MP3 files of Common Voice and Tatoeba into monochannel WAV format with a sampling rate of 16000 Hz. We then extract 80-dimensional log-mel filterbank features, using a 25ms window size and 10ms window shift. The dimension of the feature has been chosen as the best performing one among several tested. For pre-trained speech features, we use the features extracted respectively from a wav2vec model, a vq-wav2vec (kmeans) model pre-trained on Librispeech, and a BERT model pre-trained on Librispeech quantized with the corresponding vq-wav2vec model. Details of the models are provided in \autoref{sec:methods}. In the training set, samples with more than 3000 frames or having more than 400 characters are removed for GPU memory efficiency, and samples with less than 5 frames or 1 character are also removed to avoid non-significant or empty inputs. 

\subsubsection{Training and Inference}
Training and inference use the fairseq framework \cite{ott2019fairseq}. We train using the Adam optimizer \cite{kingma2014adam} with a learning rate of 1e-03 for BiLSTM models, and of 5e-05 for BERT-backbone models. We use a fixed learning schedule for BiLSTM models and a polynomial decay learning schedule for BERT-backbone models. 
In addition, we use SpecAugment \cite{park2019specaugment} for both ASR and ST with LD policy but without time warping. When training with learned features, we change the policy along the frequency dimension proportional to the embedding size. It can be thought as a kind of dropout applied to the input.

At inference time, we use beam search with a beam size of 5. We evaluate using the last 5 checkpoints averaged. For ASR, the reported word error rate (WER) has been obtained using VizSeq \cite{wang2019vizseq}. For ST, the BLEU score \cite{papineni2002bleu} reported is case-insensitive and tokenized, obtained using sacreBLEU~\cite{post2018call}.

\section{Results}

\subsection{English-to-X Speech Translation}

In this experiment, we compare the baseline log-mel filterbank features (noted as fbank) with wav2vec, vq-wav2vec and BERT features on within-language English-to-X translation, where the source audio matches the language (English) on which the learned features have been pre-trained on.

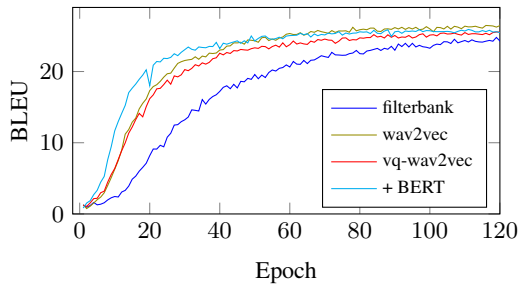
\begin{figure}[t]
    \centering
    \begin{tikzpicture}
        \begin{groupplot}[group style={group size= 1 by 1, horizontal sep=0.6cm, vertical sep=0.6 cm}, width=0.9\columnwidth, height = 0.54\columnwidth, legend cell align={left}, legend style={font=\scriptsize}]
            \nextgroupplot[ymin=0, ymax=29, xlabel=Epoch, ylabel=BLEU, y label style={yshift=-15pt}, xmin=-2, xmax=120, legend style={legend pos=south east}, xminorticks=false]
            
            \addplot[blue, mark=.] table [x index=0,y index=1, dashed, col sep=comma] {plot_mustenfr_data_dev/new_plot_fbank.csv};
            \addlegendentry{filterbank};
            
            \addplot[olive, mark=.] table [x index=0,y index=1, dashed, col sep=comma] {plot_mustenfr_data_dev/new_plot_w2v.csv};
            \addlegendentry{wav2vec};
            
            \addplot[red, mark=.] table [x index=0,y index=1, dashed, col sep=comma] {plot_mustenfr_data_dev/new_plot_vqw2v.csv};
            \addlegendentry{vq-wav2vec};
            
            \addplot[cyan, mark=.] table [x index=0,y index=1, dashed, col sep=comma] {plot_mustenfr_data_dev/new_plot_vqw2v_b.csv};
            \addlegendentry{+ BERT};
        \end{groupplot}
    \end{tikzpicture}
    \caption{Evolution of the BLEU score across epochs for different speech features on the MuST-C En-Fr dev set. The actual training has been performed until full convergence for all features.}
    \label{fig:mustcenfrbleu}
\end{figure}

\autoref{tab:res_mustc} summarizes the results obtained using different input features with the BiLSTM architecture, on the MuST-C dataset, for the English-French and English-Romanian language pairs. We can see that for both pairs, pre-trained features outperform the baseline log-mel filterbank feature. The largest improvements are obtained using the wav2vec features, with respectively 2 and 1.1 BLEU gains. Note that the MuST-C dataset is composed of TED talks (spoken English), while pre-trained features were learned on Librispeech, without need for domain adaptation. Models using pre-trained features are also found to converge faster (\autoref{fig:mustcenfrbleu}).

\begin{table}[]
  \caption{Results on the task of AST for MuST-C. The scores are computed in BLEU, on the tst-COMMON test set.}
  \label{tab:res_mustc}
  \centering
  \begin{tabular}{c|c|c}
    \toprule
    & En-Fr & En-Ro \\
    \midrule 
    Di Gangi et al.~\cite{di2019must} & 22.3 & 13.4 \\
    Di Gangi et al.~\cite{di2019adapting} & 27.9 & 16.8\\
    \midrule
    log-mel filterbank & 27.8 & 17.1\\
    wav2vec & \textbf{29.8} & \textbf{18.2} \\
    vq-wav2vec & 28.6 & 17.4 \\
    + BERT base & 28.6 & 17.3 \\
    \bottomrule
  \end{tabular}
\end{table}

\subsection{X-to-English Speech Translation}


We now investigate whether pre-trained English speech features can be transferred to other languages for the X-to-En ST task.

\subsubsection{Main Results}\label{subsubsection:xtoen}
\begin{table*}[]
    \centering
    \caption{Comparison of different speech features for English ASR and X-to-En AST. The first column indicates the WER of EN ASR models used to pre-train the ST. The ST results are on CoVoST/Tatoeba test set (when available). The ST languages are: German (De), French (Fr), Spanish (Es), Dutch (Nl), Russian (Ru), Italian (It), Turkish (Tr), Persian (Fa), Swedish (Sv), Mongolian (Mn) and Chinese (Zh). The baseline \cite{wang2020covost} is comparable to the case with ASR encoder pre-training, using log-mel filterbank features.}
        \begin{tabular}{c|c|cccccccccccc}
        \toprule
        Language & En & De & Fr & Es & Nl & Ru & It & Tr & Fa & Sv & Mn & Zh\\
        Hours (test) & & 168.3 & 46.3 & 3.5 & 8.2 & 8.2 & 12.8 & 3.8 & 23.9 & 1.0 & 2.9 & 3.7\\
        \midrule
        Wang et al. \cite{wang2020covost} & - & 7.6/7.5 & 21.4/10.9 & 6.1/1.9 & 3.4/5.0 & 4.8/1.1 & 6.5 & 3.1 & 2.8 & 1.9 & 0.3 & 5.6 \\
        \midrule
        fbank  &  & 3.1/1.5 & 17.3/6.4 & 0.8/0.5 & 0.1/0.1 & 1.3/0.1 & 0.5 & 1.1 & 0.3 & 0.2 & 0.4 & 2.4 \\
        wav2vec  & & 6/\textbf{5.0} & \textbf{21.6/12.8} & 0.4/0.4 & 0.3/0.5 & \textbf{2.0}/0.1 & 0.4 & 0.9 & 1.6 & 0.3 & 0.2 & \textbf{3.5} \\
        vq-wav2vec  & & \textbf{6.1}/5.0 & 20.8/12.2 & 0.7/0.3 & 0.2/0.4 & 2.0/0.1 & 0.5 & 0.9 & 1.2 & 0.4 & 0.3 & 3 \\
        BERT-feature & & 2.8/1.2 & 18.4/7.4 & 0.2/0.2 & 0.1/0.2 & 1.4/0.1 & 0.4 & 0.6 & 0.2 & 0.3 & 0.1 & 2.8 \\
        BERT-backbone & & 6.7 & 16.4 & 3.4 & 2.1 & 5.1 &  &  &  &  &  &  \\
        \midrule
        \multicolumn{12}{c}{With ASR encoder pre-training} \\
        \midrule
        fbank & 34.3 & 7.2/6.6 & 21.9/10.3 & 5.5/1.9 & 3.3/3.9 & 5.1/0.8 & 7.0 & \textbf{3.4} & 2.7 & 1.8 & 0.2 & \textbf{6.9} \\
        wav2vec & 32.6 & \textbf{8.6/9.7} & \textbf{22.7/14.3} & \textbf{6.5/2.4} & 3.8/5.0 & \textbf{6.1/1.3} & \textbf{8.2} & 3.4 & 3.2 & \textbf{1.9} & 0.1 & 5.8 \\
        vq-wav2vec & 35 & 8.5/9.8 & 21.9/12.4 & \textbf{6.5/2.4} & 3.7/\textbf{5.4} & 5.7/1.3 & 7.8 & 3.1 & \textbf{3.3} & 1.8 & 0.3 & 5.7 \\
        BERT-feature & 32.1 & 7.6/8.3 & 19.7/10.4 & 5.7/2.4 & \textbf{4.2}/4.2 & 5.7/1.0 & 6.6 & 3.0 & 3.1 & 1.8 & 0.3 & 5.7 \\
        \midrule
        \multicolumn{12}{c}{With ASR encoder+decoder pre-training} \\
        \midrule
        fbank &  & 8.3/7.4 & 22.5/11.2 & 6.8/2.2 & 4.0/5.5 & 8.3/1.4 & 8.8 & 3.2 & 3.1 & 3.0 & 0.2 & \textbf{8.2} \\
        wav2vec &  & \textbf{9.7}/10.1 & \textbf{23.0/14.3} & \textbf{7.2/3.6} & 4.9/6.9 & 8.8/\textbf{1.8} & \textbf{9.7} & \textbf{3.4} & \textbf{3.7} & \textbf{3.7} & 0.2 & 6.8 \\
        vq-wav2vec &  & 9.6/\textbf{11.2} & 22.1/13.1 & 6.9/3.3 & \textbf{5.0/7.0} & \textbf{9.2}/1.7 & 9.0 & 3.3 & \textbf{3.7} & 3.2 & 0.3 & 7.0 \\
        BERT-feature &  & 8.5/9.4 & 19.9/10.5 & 6.2/3.2 & 4.3/5.8 & 8.3/1.3 & 7.7 & 2.8 & 3.3 & 2.9 & 0.3 & 6.4 \\
        \bottomrule
        \end{tabular}
    \label{tab:table_covo_ast}
\end{table*}

\begin{figure}[t]
    \centering
    \begin{tikzpicture}
        \begin{groupplot}
        [group style={group size= 2 by 1, horizontal sep=0.7cm, vertical sep=0 cm}, height = 0.52\columnwidth, width=0.5\columnwidth, legend cell align={left}, legend style={font=\scriptsize}]
        \nextgroupplot[
            ylabel=BLEU, y label style={yshift=-15pt}, ymin=16, ymax=24,
            xlabel=ASR pre-training, x label style={xshift=45pt},
            symbolic x coords={None, Enc, Enc+Dec},
    xtick=data,
            legend style={legend pos=south east}, legend style={font=\scriptsize, at={(2.4,-0.65)},legend columns=4}
        ]
            \addplot[blue, mark=*] coordinates {
                (None, 17.27)
                (Enc, 21.86)
                (Enc+Dec, 22.67)
            };
            \addlegendentry{filterbank};
            \addplot[olive, mark=*] coordinates {
                (None, 21.55)
                (Enc, 22.67)
                (Enc+Dec, 23.02)
            };
            \addlegendentry{wav2vec};
            \addplot[red, mark=*] coordinates {
                (None, 20.75)
                (Enc, 21.91)
                (Enc+Dec, 22.05)
            };
            \addlegendentry{vq-wav2vec};
            \addplot[cyan, mark=*] coordinates {
                (None, 18.36)
                (Enc, 19.69)
                (Enc+Dec, 19.89)
            };
            \addlegendentry{+ BERT};
        
        \nextgroupplot[
            ymin=5,
            ymax=16,
            symbolic x coords={None, Enc, Enc-Dec},
    xtick=data,
            legend style={legend pos=south east}
        ]
            \addplot[blue, mark=*] coordinates {
                (None, 6.37)
                (Enc, 10.3)
                (Enc-Dec, 11.22)
            };
            \addplot[olive, mark=*] coordinates {
                (None, 12.81)
                (Enc, 14.27)
                (Enc-Dec, 14.33)
            };
            \addplot[red, mark=*] coordinates {
                (None, 12.23)
                (Enc, 12.40)
                (Enc-Dec, 13.05)
            };
            \addplot[cyan, mark=*] coordinates {
                (None, 7.39)
                (Enc, 10.42)
                (Enc-Dec, 10.46)
            };
        \end{groupplot}
    \end{tikzpicture}
    \caption{Comparison of BLEU scores for Fr-En ST, with/without ASR pre-training, on CoVoST test set (left) and Tatoeba test set (right)}
    \label{fig:fig_covo_fren}
\end{figure}
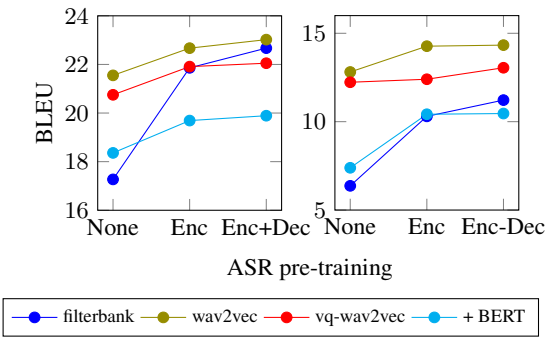
\begin{figure}[t]
    \centering
    \begin{tikzpicture}
        \begin{groupplot}
        [group style={group size= 2 by 1, horizontal sep=0.5cm, vertical sep=0 cm}, height = 0.52\columnwidth, width=0.7\columnwidth, legend cell align={left}, legend style={font=\scriptsize}]
        \nextgroupplot[ylabel=BLEU, y label style={yshift=-15pt}, xlabel=ASR pre-training, ymin=1.5, ymax=8.2,
            symbolic x coords={None, Enc, Enc+Dec},
    xtick=data, legend style={legend pos=south east}, legend style={font=\scriptsize, at={(1.6,0.39)}}]
            \addplot[blue, mark=*] coordinates {
                (None, 2.35)
                (Enc, 6.92)
                (Enc+Dec, 7.76)
            };
            \addlegendentry{filterbank};
            \addplot[olive, mark=*] coordinates {
                (None, 2.97)
                (Enc, 5.75)
                (Enc+Dec, 6.69)
            };
            \addlegendentry{wav2vec};
            \addplot[red, mark=*] coordinates {
                (None, 3)
                (Enc, 5.66)
                (Enc+Dec, 7.03)
            };
            \addlegendentry{vq-wav2vec};
            \addplot[cyan, mark=*] coordinates {
                (None, 2.83)
                (Enc, 5.66)
                (Enc+Dec, 6.42)
            };
            \addlegendentry{+ BERT};
        \end{groupplot}
    \end{tikzpicture}
    \caption{Comparison of BLEU scores for Zh-En ST, with/without ASR pre-training, on CoVoST test set (*results averaged over 4 random seeds)}
    \label{figure:cozhen}
\end{figure}
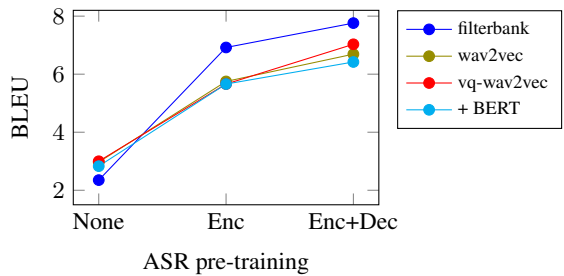

We investigate the low-resource X-to-English ST task. We consider both ST training from scratch and using an En ASR model to pre-train the ST components on the CoVoST dataset.

We report the ASR and ST results in \autoref{tab:table_covo_ast}. First, we find that while the pre-trained features are not helpful in very-low resource conditions, when there is a good baseline (either with a certain amount of data or combining with the ASR pre-training technique), they can consistently improve over the log-mel filterbank features and transfer well to other languages. On Fr-En ST, without any ASR pre-training, wav2vec features brought an improvement of 4.28/6.37 BLEU on CoVoST/Tatoeba. Second, the gain is cumulative with the ASR pre-training method to help improve low-resource ST performance, for all self-supervised features and almost all language pairs, except for Mongolian on which the systems failed to learn. Also, we observe that while on the ASR task, the most effective pre-trained feature is BERT, in the majority of X-to-En ST tasks, BERT features are outperformed by wav2vec or vq-wav2vec. 

We plot Fr-En and Zh-En results in \autoref{fig:fig_covo_fren} and \autoref{figure:cozhen} for better visualization (the general trend for most other languages is similar to French). We observe that for French, wav2vec features are consistently outperforming the baseline. In the case of Chinese, log-mel filterbank is slightly worse when we directly train the ST, but outperforms learned representations when combining with ASR pre-training.


%
We also compare the results obtained on the BERT-backbone architecture with the baseline and other self-supervised approaches, on 5 languages pairs in \autoref{tab:table_covo_ast}. The parameters transferred from the pre-trained BERT encoder can lead to better performance on 4 language pairs compared to the systems trained from scratch, but it is not as effective as using ASR pre-training. What is surprising is that the encoder contains 123.6M parameters and can still be trained effectively on low-resource setting (ex. there are only 4h of training data for Dutch).


\subsubsection{Transferring Features of Language X to English}\label{section:ft}

We now study the impact of transferring features of language X to the pre-trained speech representations or systems. We first consider directly fine-tuning a pre-trained representation. Secondly, we consider training an ASR with both source (X) and target (English) data which will then be used to warm-up the ST training.

In the first approach, we compare frozen BERT features to the features fine-tuned on Common Voice speech data (2019-06-12 release) on Fr-En and Zh-EN ST tasks. The advantage of this approach is that no labeled data is required. \autoref{tab:covo_ft_bert} shows that fine-tuning is helpful in all cases, except for Zh-En ST without ASR pre-training. On both language pairs, combining fine-tuned features with ASR pre-training is more helpful when pre-training only the encoder. 

\begin{table}[t]
    \centering
    \caption{BLEU scores using BERT features fine-tuned on language X. The difference compared to the frozen features (row \emph{BERT-feature} in Table 3) is in parentheses.}
    \begin{tabular}{c|cc}
        \toprule
        ASR pre-training & Fr & Zh \\
        \midrule
        None & 18.7 (+0.3) & 2.0 (-0.8)\\
        Encoder & 21.0 (+1.3) & 6.8 (+1.1) \\
        Encoder+Decoder & 20.9 (+1.0) & 6.7 (+0.3) \\
        \bottomrule
    \end{tabular}
    \label{tab:covo_ft_bert}
\end{table}

\begin{table}[t]
    \caption{WER for En+X ASR and BLEU for the corresponding ST, using encoder pre-training. Difference with respect to En ASR is in parentheses: for ASR, it is computed against the 1st column of Table 3, for AST against the respective languages of Table 3 for the encoder pre-training case. A, B, C and D refer to fbank, wav2vec, vq-wav2vec and BERT features, respectively.}
    \centering
    \begin{tabular}{c|cccc}
    \toprule
     & De & Fr & Es & Zh \\
     \midrule
     \multicolumn{5}{c}{ASR} \\ 
     \midrule
     A & 35.9 (+1.6) & 34.7 (+0.4) & 34.7 (+0.4) & 37.2 (+2.9) \\
     B & 33.5 (+0.9) & 32.1 (-0.5) & 32.9 (+0.3) & 36.0 (+3.4) \\
     C & 35.4 (+0.4) & 34.7 (-0.3) & 35.9 (+0.9) & 37.7 (+2.7) \\
     D & 35.0 (+2.9) & 32.7 (+0.6) & 32.8 (+0.7) & 33.2 (+1.1) \\
     \midrule
     \multicolumn{5}{c}{AST} \\
     \midrule
     A & 8.3 (+1.1) & 23.2 (+1.3) & 7.4 (+1.9) & 7.5 (+0.6)\\
     B & 9.3 (+0.7) & 23.9 (+1.2) & 8.4 (+1.9) & 7.3 (+1.5)\\
     C & 9.5 (+1.0) & 22.8 (+0.9) & 7.7 (+1.2) & 7.2 (+1.5) \\
     D & 8.4 (+0.8) & 20.9 (+1.2) & 7.8 (+2.0) & 7.2 (+1.5) \\
    \bottomrule
    \end{tabular}
    \label{tab:table_wer_bleu}
\end{table}

In the second approach, we leverage ASR data and investigate the impact of mixing source language X with English data to train the ASR model which will then be used to fine-tune the encoder of the ST model. For both English and X, we use the Common Voice ASR training data. \autoref{tab:table_wer_bleu} shows the results for 4 language pairs from higher-resource to low-resource settings. While combining different languages may increase the WER of the ASR, it can still help improve the performance of the resulting ST in all cases. Also, for most languages, pre-trained representations can also improve over the baseline log-mel filterbank in this setting.

We observe that on Fr-En and Es-En ST, for all the 4 features, pre-training only the ST encoder with the En+X ASR is performing even better than pre-training both ST encoder and decoder with the En ASR (in \autoref{tab:table_covo_ast}). The largest gaps have been observed on BERT features, with respectively a difference of 1 and 1.6 BLEU for Fr-En and Es-En.

\subsubsection{Influence of ASR Performance}

The experiments in sections \ref{subsubsection:xtoen} and \ref{section:ft} suggest that when the training conditions differ, i.e. when comparing ASR models pre-trained on different features and/or on different languages, the ASR WER may not necessarily be correlated with the performance of the final AST.

\autoref{tab:table_covo_ast} (column En) shows that while vq-wav2vec led to the worst performance on En ASR, in most cases, the final ST results are better than the systems pre-trained on En ASR with BERT features, whose WER is 2.9 points lower.

This effect is even more pronounced in \autoref{tab:table_wer_bleu}, where in most cases, ASR models with higher WER can still help improve the translation performances.

\section{Conclusion}

We have shown that self-supervised representations can benefit the ST task. The resulting features can be directly transferred to other languages, and can be effectively combined with ASR pre-training for low-resource conditions to boost the performance. To improve the cross-lingual transfer on a given language, an effective way is to leverage ASR data by transferring the parameters learned on an ASR pre-trained on both higher-resource English and X data, or fine-tuning the pre-trained features on language X in an unsupervised way. Further work can include analyzing investigating the robustness of pre-trained features in other data conditions, and exploring multilingual settings.


\bibliographystyle{IEEEtran}

\bibliography{mybib}

\end{document}